\newif{\ifremarks}

\remarkstrue  

\ifremarks
\newcommand{\remarktb}[1]{{\renewcommand{\bfdefault}{b}{\color[RGB]{0,150,0}{\textbf{#1}}}}}
\newcommand{\remarkjc}[1]{{\renewcommand{\bfdefault}{b}{\color[RGB]{0,0,150}{\textbf{#1}}}}}
\newcommand{\remarkpv}[1]{{\renewcommand{\bfdefault}{b}{\color[RGB]{150,0,0}{\textbf{#1}}}}}
\fi
\providecommand{\remarktb}[1]{\ignorespaces}
\providecommand{\remarkjc}[1]{\ignorespaces}
\providecommand{\remarkpv}[1]{\ignorespaces}

\documentclass[
 reprint,
 amsmath,
 amssymb,
 aps,
 prl,
 longbibliography,
 superscriptaddress
]{revtex4-1}

\usepackage[utf8]{inputenc}

\usepackage[pdftex]{graphicx}
\usepackage{wasysym}
\usepackage{amsfonts}
\usepackage{ifsym}

\makeatletter
\newlength{\apb@width}
\newcommand{\autoparbox}[2][c]{\settowidth{\apb@width}{#2}\parbox[#1]{\apb@width}{#2}}
\newcommand{\includegraphicsbox}[2][]{\autoparbox{\includegraphics[#1]{#2}}}
\makeatother

\newcommand{\namedref}[2]{\hyperref[#2]{#1~\ref*{#2}}}

\newcommand{\tabref}[1]{\namedref{Table}{#1}}
\newcommand{\figref}[1]{\namedref{Figure}{#1}}

\allowdisplaybreaks
\def\etal.{et\penalty50\ al.}
\usepackage{xspace}
\newcommand*{\eg}{e.\,g.\@\xspace}
\newcommand*{\ie}{i.\,e.\@\xspace}
\makeatletter\newcommand*{\etc}{%
    \@ifnextchar{.}%
        {etc}%
        {etc.\@\xspace}%
}\makeatother

\let\oldbfseries=\bfseries
\let\oldmdseries=\mdseries
\let\oldnormalfont=\normalfont
\renewcommand{\bfseries}{\oldbfseries\boldmath}
\renewcommand{\mdseries}{\oldmdseries\unboldmath}
\renewcommand{\normalfont}{\oldnormalfont\unboldmath}

\newcommand{\suprm}[1]{^{\text{#1}}}
\newcommand{\subrm}[1]{_{\text{#1}}}

\newcommand{\grp}[1]{\mathrm{#1}}

\newcommand{\mathematica}{\textsc{Mathematica}}

\newcommand{\brk}[1]{(#1)}

\newcommand{\bigbrk}[1]{\bigl(#1\bigr)}

\newcommand{\sbrk}[1]{[#1]}

\newcommand{\Bigsbrk}[1]{\Bigl[#1\Bigr]}
\newcommand{\biggsbrk}[1]{\biggl[#1\biggr]}

\newcommand{\biggbrc}[1]{\biggl\{#1\biggr\}}

\newcommand{\vev}[1]{\langle#1\rangle}

\newcommand{\tr}{\operatorname{tr}}

\newcommand{\op}[1]{\mathcal{#1}}

\newcommand{\order}[1]{\mathcal{O}(#1)}

\newcommand{\superN}{\mathcal{N}}
\newcommand{\gym}{g_{\scriptscriptstyle\mathrm{YM}}}

\newcommand{\dd}{\mathrm{d}}
\newcommand{\Nc}{N\subrm{c}}
\newcommand{\Csphere}{{}^\bullet\kern-1.2pt C}
\newcommand{\Ctorus}{{}^\circ\kern-1.2pt C}

\newcommand{\nn}{\nonumber}

\newcommand{\COMMENT}[1]{}

\newcommand{\neqa}{\nonumber\end{eqnarray}}

\newcommand{\<}{{\langle}}
\renewcommand{\>}{{\rangle}}

\newcommand{\re}{\relax{\rm I\kern-.18em R}}

\def\su2{{SU(2)}}

\def\[{\left[}
\def\]{\right]}

\def\({\left(}
\def\){\right)}
\def\[{\left[}
\def\]{\right]}

\def\<{\langle}
\def\>{\rangle}

\def\i2{\frac{i}{2}}

\def\2F1{\,_2{\rm F}_1}

\usepackage[usenames,dvipsnames]{xcolor}
\usepackage{color}
\usepackage{graphicx}
\usepackage{dcolumn}
\usepackage{bm}
\usepackage{hyperref}

\newcommand{\beq}{\begin{equation}}
\newcommand{\eeq}{\end{equation}}
\newcommand{\beqq}{\begin{equation*}}
\newcommand{\eeqq}{\end{equation*}}
\newcommand\beqa{\begin{eqnarray}}
\newcommand\eeqa{\end{eqnarray}}
\newcommand\beqaa{\begin{eqnarray*}}
\newcommand\eeqaa{\end{eqnarray*}}
\newcommand\bea{\begin{array}}
\newcommand\eea{\end{array}}

\newcommand{\MZ}{Minahan:2002ve}
\newcommand{\BPR}{Bena:2003wd}
\newcommand{\FT}{Berkovits:2008ic,Beisert:2008iq}
\newcommand{\BigReview}{Beisert:2010jr}
\newcommand{\BKS}{Beisert:2003tq}

\newcommand{\PentagonPaper}{Basso:2013vsa}
\newcommand{\HexagonPaper}{Basso:2015zoa}
\newcommand{\Hexagonalizationone}{Fleury:2016ykk}
\newcommand{\Cushions}{Eden:2016xvg}
\newcommand{\Hexagonalizationtwo}{Fleury:2017eph}
\newcommand{\ShotaFriends}{Kim:2017phs}
\newcommand{\OriginalTwistOpsPaper}{Cardy:2007mb}
\newcommand{\VeryRecentTwistOpsPaper}{Castro-Alvaredo:2017wzf}
\newcommand{\AdSLoops}{Rastelli:2016nze,Aharony:2016dwx,Alday:2017xua,Aprile:2017bgs,Aprile:2017xsp,Rastelli:2017udc,Alday:2017vkk,Aprile:2017qoy}

\begin{document}

\title{Handling Handles: Nonplanar Integrability \texorpdfstring{\\}{} in
\texorpdfstring{$\superN=4$}{N=4} Supersymmetric Yang--Mills Theory}

\author{Till Bargheer}
\affiliation{Institut f\"ur Theoretische Physik, Leibniz Universit\"at Hannover, Appelstra{\ss}e 2, 30167 Hannover, Germany}
\affiliation{DESY Theory Group, DESY Hamburg, Notkestra{\ss}e 85, D-22603 Hamburg, Germany}
\affiliation{Kavli Institute for Theoretical Physics, University of California, Santa Barbara, CA 93106, USA}
\author{Jo\~ao Caetano}
\affiliation{Laboratoire de Physique Th\'eorique, \'Ecole Normale Sup\'erieure \& PSL Research University, 24 rue Lhomond, 75231 Paris Cedex 05, France}
\author{Thiago Fleury}
\affiliation{Laboratoire de Physique Th\'eorique, \'Ecole Normale Sup\'erieure \& PSL Research University, 24 rue Lhomond, 75231 Paris Cedex 05, France}
\affiliation{Instituto de F\'isica Te\'orica, UNESP, ICTP South American Institute for Fundamental Research, Rua Dr Bento Teobaldo Ferraz 271, 01140-070, S\~ao Paulo, Brazil}
\affiliation{International Institute of Physics, Federal University of Rio Grande do Norte, Campos Universitario, Lagoa Nova, Natal, Rio Grande do Norte 59078-970, Brazil}
\author{Shota Komatsu}
\affiliation{School of Natural Sciences, Institute for Advanced Study, Einstein Drive, Princeton, NJ 08540, USA}
\affiliation{Perimeter Institute for Theoretical Physics, 31 Caroline Street North, Waterloo, Ontario N2L 2Y5, Canada}
\author{Pedro Vieira}
\affiliation{Instituto de F\'isica Te\'orica, UNESP, ICTP South American Institute for Fundamental Research, Rua Dr Bento Teobaldo Ferraz 271, 01140-070, S\~ao Paulo, Brazil}
\affiliation{Perimeter Institute for Theoretical Physics, 31 Caroline Street North, Waterloo, Ontario N2L 2Y5, Canada}

\begin{abstract}
We propose an integrability setup for the computation of correlation functions of gauge-invariant
operators in $\mathcal{N}=4$ supersymmetric Yang--Mills theory at higher orders in the
large $\Nc$ genus expansion and at any order in the 't Hooft
coupling $\gym^2\Nc$.
In this multi-step proposal, one polygonizes the string worldsheet in
all possible ways, hexagonalizes all resulting polygons, and sprinkles mirror particles over all hexagon
junctions to obtain the full correlator.
We test our integrability-based conjecture against a non-planar four-point correlator of large half-BPS
operators at one and two loops.
\end{abstract}

\maketitle

\section{Introduction}

\textit{Integrable theories} are rather special 2D quantum field theories where the
scattering of fundamental excitations factorizes into a
sequence of two-body scattering events. This
simplification often translates into solvability. The worldsheet theory describing superstrings in $\grp{AdS}_5\times\grp{S}^5$ is
integrable~\cite{\MZ,\BPR}.
Exploiting integrability machinery, the full \textit{finite-size
spectrum} has been
obtained at any value of the coupling~\cite{\BigReview,Gromov:2013pga,Gromov:2014caa},
yielding the energy spectra of single strings in this
curved background or -- equivalently -- the spectra of anomalous
dimensions of single-trace operators in $\mathcal{N}=4$ supersymmetric
Yang--Mills (SYM) theory in the planar limit.

\begin{figure}[t]
\centering
\includegraphics[scale=0.30]{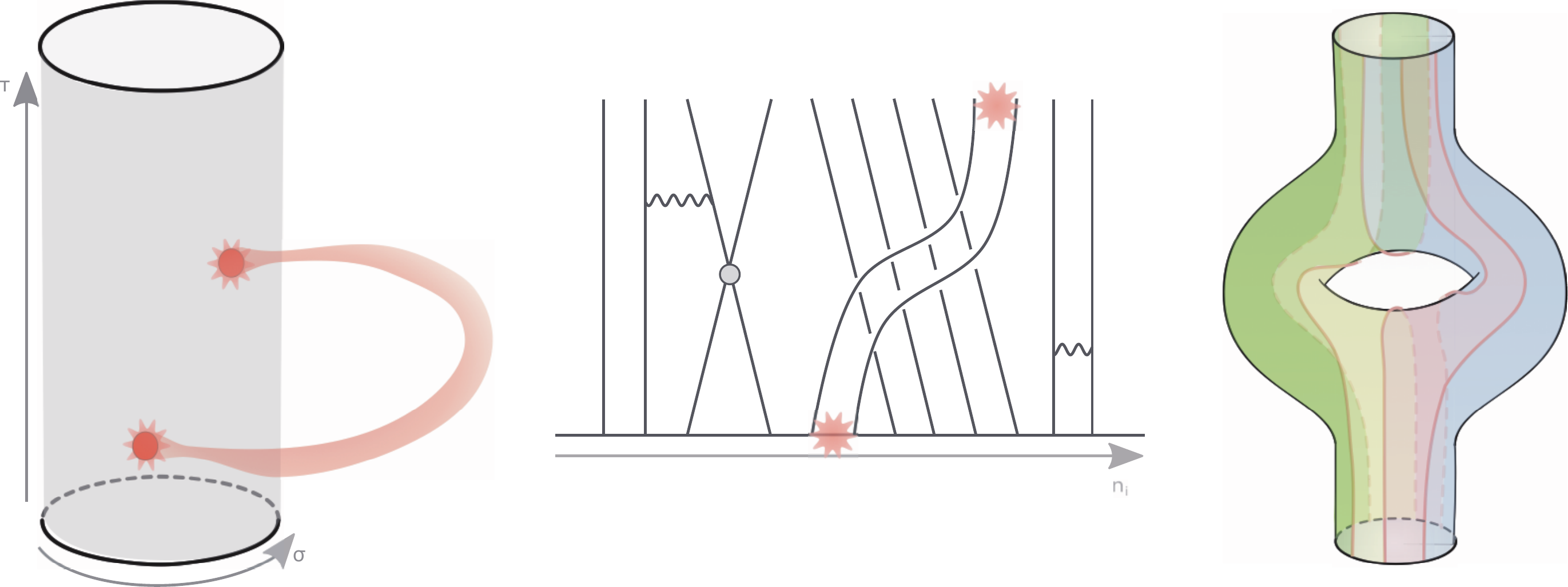}
\caption{(a) Non-planar effects include handles in the string
world-sheet inducing nontrivial non-local effects. (b) The same effect
can be seen in the gauge theory; non-planar processes induce
non-local interactions in the effective spin chain. (c) From a
hexagonalization point of view, we tessellate higher-genus string
topologies, always maintaining locality.\label{fig:handles}}
\end{figure}

Beyond the planar limit, we are dealing with worldsheets with handles.
These induce non-local
interactions in the two-dimensional theory, wormholes of sorts, which also appear in the gauge
theory spin-chain description, see \figref{fig:handles}.
One would guess that such
non-local interactions could ruin integrability. Indeed, known
degeneracies in the spectrum of the weakly coupled gauge theory --
related to the hidden higher charges of the integrable theory -- are lifted as one takes non-planar corrections
into account~\cite{\BKS}, and fermionic T-duality --
responsible for dual conformal symmetry, which in turn is closely
related to integrability in the usual sense -- is not a symmetry of
string theory at higher genus~\cite{\FT}. Because of all this, it has
been common lore that integrability would not be useful beyond the
planar limit
\footnote{For operators whose dimension scale as $\Delta \sim \order{\Nc}$,
integrability at the non-planar level was discussed previously
in~\cite{Carlson:2011hy,Koch:2011jk,deMelloKoch:2012sv}}.
See~\cite{Kristjansen:2010kg} for a very nice summary.

On the other hand, numerous other \emph{planar} quantities have been
explored at finite coupling using integrability, from
scattering amplitudes or Wilson loops~\cite{\PentagonPaper} to
structure constants~\cite{\HexagonPaper},
higher-point correlation
functions~\cite{\Hexagonalizationone,\Cushions,\Hexagonalizationtwo},
and even
mixed quantities involving correlation functions in
the presence of Wilson loops~\cite{\ShotaFriends}.
Underlying all these computations is the idea of taming complicated
string topologies by cutting the string into smaller and
simpler patches (hexagonal or pentagonal), which are then glued back
together. This is implemented by so-called branch-point
twist field
operators~\cite{\OriginalTwistOpsPaper,\VeryRecentTwistOpsPaper}, whose
expectation values can be bootstrapped.

All these works strongly suggest that, instead of thinking about the non-planar effects as
non-local corrections to the planar world-sheet, we should from the
get-go consider the theory in more general
topologies,
treat handles using the
twist operators mentioned above, and keep
everything else as local as possible, see \figref{fig:handles}.
Following this philosophy, in this Letter, we propose a framework for
computing correlation functions at any higher-genus order and any
value of the 't~Hooft coupling using integrability.

\section{The data}

Our \textit{experimental data} -- against which we will test our integrability predictions -- are the four-point correlation functions of single-trace half-BPS
operators~$\op{Q}_i^k\equiv \tr\brk{\sbrk{\alpha_i{\,\cdot\,}\Phi(x_i)}^k}$ studied
in~\cite{Arutyunov:2003ae,Arutyunov:2003ad} in the simplifying
configuration~$\alpha_1{\cdot}\alpha_4=\alpha_2{\cdot}\alpha_3=0$.
Here, $\alpha_i$ is a null vector, and $\Phi=(\phi_1,\dots,\phi_6)$
are the scalar fields of $\superN=4$ SYM theory.
The loop correlator $G_k \equiv
\vev{\op{ Q}^k_1\op{Q}^k_2\op{Q}^k_3\op{ Q}^k_4}
-\vev{\op{  Q}^k_1\op{Q}^k_2\op{Q}^k_3\op{  Q}^k_4}\suprm{tree}$ can then be decomposed
according to the propagator structures that connect the operators as
\begin{equation}
G_k=\sum_{m=0}^{k}\left(\sum_{l=1}^\infty g^{2l} \mathcal{F}^{(l)}_{k,m}\right)\,X^mY^{k-m},
\label{eq:corr}
\end{equation}
where $X\equiv
(\alpha_1{\,\cdot\,}\alpha_2)(\alpha_3{\,\cdot\,}\alpha_4)/x_{12}^2 x_{34}^2$,
$Y\equiv X |_{2\leftrightarrow 3}$ are the R-charge and space-time propagators,
and~$g^2 = g\subrm{YM}^2\Nc/16\pi^2$.
The quantum corrections
dressing the propagator structures depend
on the conformally invariant cross ratios
$|z|^2=x_{12}^2x_{34}^2/x_{13}^2x_{24}^2$ and $|1-z|^2=
x_{23}^2x_{14}^2/x_{13}^2x_{24}^2$.
The one- and two-loop contributions
were computed
in~\cite{Arutyunov:2003ae,Arutyunov:2003ad}. A key ingredient
are the conformal box and double-box functions
\begin{align}
&F^{(1)}(z,\bar z)
=
\frac{x_{13}^2x_{24}^2}{\pi^2}
\int\frac{\dd^4x_5}{x_{15}^2x_{25}^2x_{35}^2x_{45}^2}
=\includegraphicsbox{FigBoxInt.mps}
\,,
\label{eq:box}
\\
&\frac{F^{(2)}}{x^2_{14}}
=
\frac{x_{13}^2x_{24}^2}{(\pi^2)^2}
\!\int\!\frac{\dd^4x_5\,\dd^4x_6}{x_{15}^2x_{25}^2x_{45}^2x_{56}^2x_{16}^2x_{36}^2x_{46}^2}
=\includegraphicsbox{FigDoubleBoxInt.mps}
.
\nonumber
\end{align}
Other key players are the so-called color factors, which
consist of color contractions of four symmetrized traces from the four
operators,
dressed with insertions of gauge group structure constants.
For instance
\footnote{Here, $\tr\brk{\brk{a_1\dots a_k}}\equiv\tr\brk{T^{(a_1}\dots
T^{a_k)}}$ denotes a totally symmetrized trace of adjoint gauge group
generators $T^a$},
\begin{align}
  C\suprm{c}_m&=\includegraphicsbox{FigCc.mps}
  =\frac{f_{abe}f_{cd}{}^ef_{pqt}f_{rs}{}^t}{2m!^2(k-m-2)!^2} \times\\
&\times  \tr \brk{\brk{d_1 \,...\, d_{k'}a_1 \,...\,  a_mbd}}
     \tr \brk{\brk{a_1 \,...\,  a_mb_1 \,...\,  b_{k'}ar}}
     \nn\\
&     \times\,
     \tr \brk{\brk{d_1 \,...\,  d_{k'}c_1 \,...\,  c_mcp}}
     \tr \brk{\brk{c_1 \,...\,  c_mb_1 \,...\,  b_{k'}qs}}
     \,,\nn
\end{align}
where $k'=k-m-2$.
We explicitly performed the contractions with \mathematica, for up to
$k=8$ or~$9$ and various values of $m$.
Then, we used the fact that -- by their combinatorial nature -- the
various color factors should be quartic polynomials in $k$ and
$m$ (up to boundary cases at extremal values
of $k$ or $m$), which we can fit using the data points at finite $k$
and $m$. At the end of the day, one finds $C_m^c/\Nc^{2k}
k^4=2k^4+\mathcal{P}^c/6\Nc^2 + \mathcal{O}(\Nc^{-4})$,
where
$\mathcal{P}^c=
k^4 + 4 k^3 m + 42 k^2 m^2
- 92 k m^3
+ 46 m^4 +\dots
$,
and similar expressions for all other color factors.

We consider the further simplification of large external operators
with $k\gg 1$ and $m/k \equiv r+1/2$ held fixed. Putting the above
ingredients together, and keeping only the leading large $k$
result at each genus order, we finally obtain our much desired
experimental data
\begin{align}
&\mathcal{F}_{k,m}^{(1)}
=
\frac{-2 k^2}{
\Nc^2}\biggbrc{
1+\frac{k^4\brk{\frac{17 r^4}{6}-
\frac{7 r^2}{4}+\frac{11}{32}
}
}{\Nc^2}}
|z-1|^2F^{(1)}
\,,
\nn\\
&\mathcal{F}_{k,m}^{(2)}
=
\frac{4 k^2}{
\Nc^2}
\biggsbrk{\biggbrc{
1+\frac{k^4
(\frac{17 r^4}{6}-\frac{7}{4} r^2+\frac{11}{32})}{ \Nc^2}
}
|z-1|^2 F^{(2)}
\nn\\
&\,\,\, +
\biggbrc{
1+\frac{k^4 (\frac{29 r^4}{6}-\frac{11}{4} r^2+\frac{15}{32})}{\Nc^2}
}
\frac{|z-1|^4}{4}
\bigbrk{F^{(1)}}^2
}
\,.
\label{eq:Fkm2largek}
\end{align}

\section{Integrability Proposal}

\begin{figure}[t]
\centering
\includegraphics[width=\columnwidth]{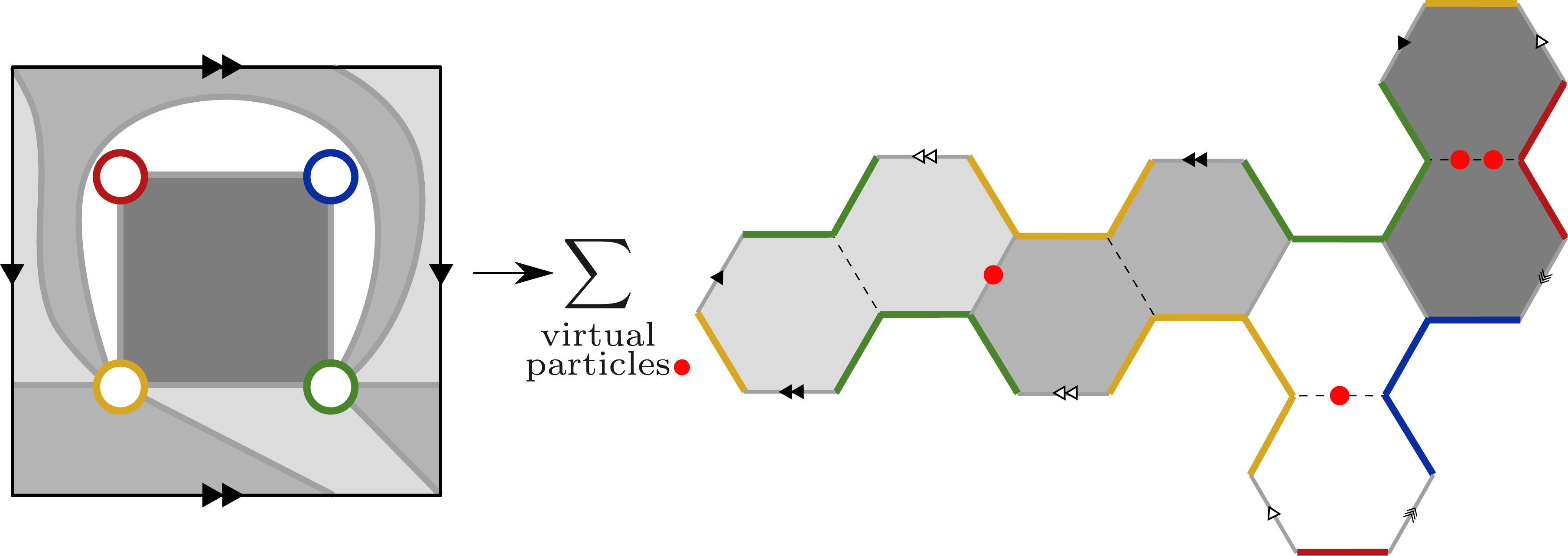}
\caption{We sum over all polygonizations of the torus with four operators/holes
(left figure). Each polygonization is then broken apart into
hexagons (right figure, edges with the same arrow marks are identified).
Finally, we dress the
hexagon junctions
with mirror particles (sprinkling). In this example, we sprinkle $1+1+2$
mirror particles on one such hexagonalization.
The two particles on a zero-length bridge (right) and
the single particle on a bridge of non-zero length $l$ (left)
kick in at $4$ and at $l+1$ loops respectively, and are thus highly
suppressed (for large bridges). The remaining (middle) contribution
with a single particle on a zero-length bridge is the only one
relevant for this note; it kicks in at one loop
already.\label{fig:3steps}}
\end{figure}

We propose that the connected part of any correlator in the
$\grp{U}(\Nc)$ theory, including the
full expansion in $1/\Nc$, can be recovered from integrability via the
formula
\begin{multline}
\vev{\op{Q}_1\dots\op{Q}_n}
={\color{magenta}\mathcal{S}\circ}
\!\label{eq:master}\!
{\color{blue}\sum_{\substack{\text{skeleton}\\\text{graphs}}}
\sum_{\substack{\text{labelings}}}
\sum_{\substack{\text{bridge}\\\text{fillings}}}
\text{\footnotesize $\! \left(\! \begin{array}{c}
\texttt{R-charge \&} \\
\texttt{space-time} \\
\texttt{propagators}
\end{array}\!\right)$}
\times}
\\\qquad\times
{\color{ForestGreen}\sum_{\substack{\text{mirror}\\\text{states}}}
\text{\footnotesize $\! \left(\! \begin{array}{c}
\texttt{mirror}\\
\texttt{propagation}\\
\texttt{factors}
\end{array}\!\right)$}
\!\times\!
{\color{red} \prod\limits_{\text{Hexagons}}\text{\footnotesize $
\left(\!\begin{array}{c}
\texttt{hexagon}\\
\texttt{form}\\
\texttt{factors}
\end{array}\!\right)$}}}.
\end{multline}
The outermost sum runs over
all graphs with $n$ vertices, including all topologies, planar and
non-planar.
Each edge (bridge) stands for a collection of one or more
(planar, non-crossing) propagators connecting two operators. Hence
parallel edges must be identified, and this defines a skeleton graph, see~\figref{fig:maxcycgraphs} for examples of such graphs.

Next, we sum over all vertex
labelings (distributions of operators on the vertices), and over all
(nonzero) bridge fillings (numbers of propagators on each edge)
compatible with the charges of the operators.
All this combinatorial process is
what we call
{\color{blue}\emph{{polygonization}}}.

Next follows what we call
{\color{red}\emph{hexagonalization}}:
After inserting the operators, all faces of the skeleton graphs are
hexagons or
higher polygons. For the latter, we pick a subdivision into hexagons by
inserting
zero-length bridges (ZLBs).
Each hexagon gives home to one
hexagon form factor whose expression was determined in~\cite{\HexagonPaper}.
Finally, we cut the graphs at the zero-length
and non-zero-length bridges, and we insert a complete basis of
mirror states, \ie we sum over mirror
excitations, on each bridge.
The mirror propagation factors depend
on the normalization and flavor of the
mirror-particle states as well as on the bridge length;
their expressions can be found in~\cite{\HexagonPaper,\Hexagonalizationone}. This last step
we denote as
{\color{ForestGreen}\textit{sprinkling}}.

These three main processes
are represented in~\figref{fig:3steps} and discussed in detail below.
For illustration and simplicity, in this Letter, we restrict
ourselves to $n=4$ large BPS operators computed up to the first
subleading correction in $1/\Nc^2$ (\ie genus 0 and genus 1).

There is one last step represented by the seemingly
innocuous $\mathcal{S}$ in~\eqref{eq:master} which stands for
subtractions or
{\color{magenta}\textit{stratification}}.
The point is that the sum over polygonizations discretizes the
integration over the moduli space
of the Riemann surface, whose boundary
contains degeneration points: At its boundary, a torus degenerates
into a sphere for instance.
$\mathcal{S}$ stands for the appropriate subtractions which remove
these boundary contributions, see \eg~\cite{Chekhov:1995cq}.
In this Letter,
we will consider four large BPS operators on the torus, which are
controlled by configurations where all cycles of the torus will be
populated by many propagators. The relevant worldsheets are thus very
far from the boundary of the moduli space, and we
can ignore $\mathcal{S}$ altogether. We will come back to it
in~\cite{Bargheer:2018jvq}, but the essential idea is that to obtain the
correct result at a given genus $g$, we must include the contributions
of graphs with genus smaller than $g$ embedded on a genus $g$ surface,
and subtract all the degenerations of that surface that do not affect
the embedded graph.

\section{(Large \texorpdfstring{$k$}{k}) Polygonization}

\begin{figure}[t]
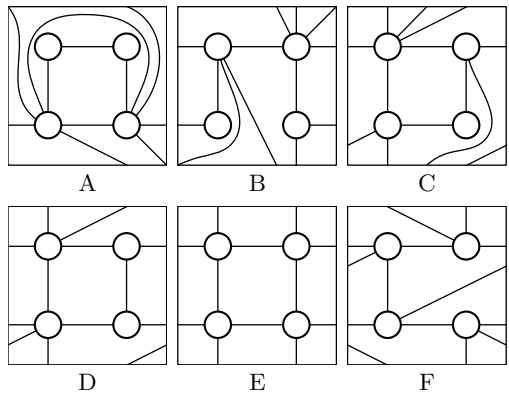

\centering
\begin{tabular}{ccc}
\includegraphicsbox{FigTorusCaseF.mps}
& \includegraphicsbox{FigTorusCaseB.mps}
& \includegraphicsbox{FigTorusCaseI.mps}
\\
\rule{0pt}{2.5ex}%
A & B & C
\\
\rule{0pt}{9ex}%
\includegraphicsbox{FigTorusCaseJ.mps}
& \includegraphicsbox{FigTorusCaseL.mps}
& \includegraphicsbox{FigTorusCaseM.mps}
\\
\rule{0pt}{2.5ex}%
D & E & F
\end{tabular}
\caption{Bridge configurations on the torus that contribute to the
leading term in $1/k$ for correlators of the
type~\eqref{eq:corr}.\label{fig:maxcycgraphs}}
\end{figure}

As indicated by the first line of~\eqref{eq:master},
the polygonization proceeds in three steps: (A) construct all
inequivalent graphs with $n$ vertices on the given topology, (B) sum
over all inequivalent
labelings of the vertices, and (C) for each labeled graph, sum over all possible
distributions of propagators on the edges (bridges) of the graph, such
that each edge carries at least one propagator.

In a generic graph on the torus, any two operators will be connected
by one or more bridges.
In this work, we are interested in the leading contribution for large
operator weights $k\gg1$ with $m/k$ finite.
In this limit, graphs with a non-maximal number of bridges will
be suppressed by combinatorial powers of $1/k$. Namely, distributing
$n\sim k$ propagators on $j$ bridges comes with a factor
\begin{equation}
\sum_{\substack{1\leq n_1,\dots,n_j\leq n\\\sum_in_i=n}}1
=\frac{n^{j-1}}{(j-1)!}
+\order{n^{j-2}}
\,.
\label{eq:lengthsum}
\end{equation}
In the leading term, all bridges carry many propagators.
We consider operator polarizations that disallow
propagator structures of the type
$Z\equiv(\alpha_1\cdot\alpha_4)(\alpha_2\cdot\alpha_3)/x_{14}^2x_{23}^2$,
see~\eqref{eq:corr}. Hence only graphs where the four operators are
connected cyclically, as in \mbox{1--2--3--4--1}, will contribute.
Under this constraint, one easily finds that the maximal power from
combinatorial factors~\eqref{eq:lengthsum} is $k^4$. We have
classified all graphs contributing to this order,
and have found the six cases shown in~\figref{fig:maxcycgraphs}.

For these six graphs, we have to consider all possible inequivalent operator
labelings. In addition, each labeled
graph comes with a combinatorial factor~\eqref{eq:lengthsum}.
We list all inequivalent labelings for the relevant graphs as well as
their combinatorial factors in~\tabref{tab:labeltab}
\begin{table}
\caption{All inequivalent operator labelings for the graphs that
contribute to leading order in $1/k$, together with their
combinatorial factors according to~\eqref{eq:lengthsum}. The order of
the labels runs from top to bottom, left to right in the graphs
of~\figref{fig:maxcycgraphs}.\label{tab:labeltab}}
\begin{ruledtabular}
\begin{tabular}{ccc}
Case & Inequivalent Labelings & Combinatorial Factor\\
\colrule
A &  $1234, 3412$ & $m^4/24$ \\
A &  $1324, 2413$ & $(k-m)^4/24$ \\
B &  $1234, 2143, 3412, 4321$ & $m^3(k-m)/6$ \\
B &  $1324, 3142, 2413, 4231$ & $m(k-m)^3/6$ \\
C &  $1234, 3412, 2143, 4321$ & $m^2/2\cdot(k-m)^2/2$ \\
D &  $1234, 2143, 1324, 3142$ & $m^2(k-m)^2/2$ \\
E &  $1234$ & $m^2(k-m)^2/2$ \\
F &  $1234$ & $m^2(k-m)^2$ \\
\end{tabular}
\end{ruledtabular}
\end{table}
\footnote{Case E has
an extra symmetry: Every pair of operators is connected by a pair of
bridges. Exchanging the members of all pairs simultaneously amounts to
a cyclic rotation of the four operators and thus leaves the
configuration invariant. The naive sum over bridge lengths thus has to
be corrected by a factor of $1/2$}.

\section{(Large \texorpdfstring{$k$}{k}) Hexagonalization}

Next, we further decompose all polygons
in~\figref{fig:maxcycgraphs} -- which are bounded by the finite bridges
-- into hexagons by adding ZLBs. There are typically
various ways of adding these ZLBs, and they are all equivalent. The
independence on the tessellation (chosen set of additional
ZLBs) was verified explicitly in the case of the octagon and decagon
in~\cite{\Hexagonalizationone,\Hexagonalizationtwo}, and represents a
consistency check of the hexagonalization.
We can easily see that all
graphs in~\figref{fig:maxcycgraphs} are made out of four
octagons; hence, we simply need to split each of those octagons into two
hexagons.
A hexagonalization of case~A is
illustrated in~\figref{fig:3steps}.
The physical operators correspond to the thick colorful lines, the solid gray
lines are the large bridges, and the dashed lines are the ZLBs.

\section{(Large \texorpdfstring{$k$}{k}) Sprinkling}

Finally, we have to sprinkle mirror particles on the hexagonalizations of the
previous section. Our large $k$ result is given by a set of octagons
separated by large bridges. Putting particles on those bridges is very
costly in perturbation theory, as the mirror-particle contribution is coupling-suppressed by
the corresponding bridge length.
Hence, we can only put
particles on ZLBs inside each octagon. Furthermore, putting two particles on
the same bridge is also very costly (appearing at four loops only), so
up to two loops only two contributions will matter: A single particle
placed on a ZLB, and two particles placed simultaneously on two
distinct ZLBs.
This latter contribution is essentially the square of the former one.
The single-particle contribution has been studied
in~\cite{Fleury:2016ykk}
and yields
\begin{equation}
\!\!\!\mathcal{M}^{(1)} =  \Bigsbrk{ z+\bar{z}-\left(\alpha +
\bar{\alpha}\right) \frac{\alpha  \bar{\alpha}+z \bar{z}}{2 \alpha
\bar{\alpha}}} \big(g^2 F^{(1)}-2 g^4 F^{(2)}\big)\,,
\label{eq:voila}
\end{equation}
where for the correlators considered here,
the R-charge cross-ratios $\alpha$ and
$\bar \alpha$ are given by
\begin{equation}
\label{eq:crpolarization}
\alpha = z \bar{z}\,{X}/{Y}
\,,\quad \text{and}\quad
\bar{\alpha} = 1 \,.
\end{equation}
To get the $g^4$ term in~\eqref{eq:voila}, we simply expanded the integrand
in~\cite{Fleury:2016ykk} to one more order in perturbation theory.

The above factors of $X$ and $Y$ are contained in the factors in the
first (green)
parentheses in the second line of~\eqref{eq:master}, and combine with
the propagator factors in the first line of that formula. Hence to
read off particular coefficients of monomials in $X$ and $Y$ to
compare with perturbation theory predictions such as~\eqref{eq:Fkm2largek},
we often need to consider the contribution of a
few ``neighboring'' graphs.

\begin{figure}
\centering
\includegraphics[scale=0.3]{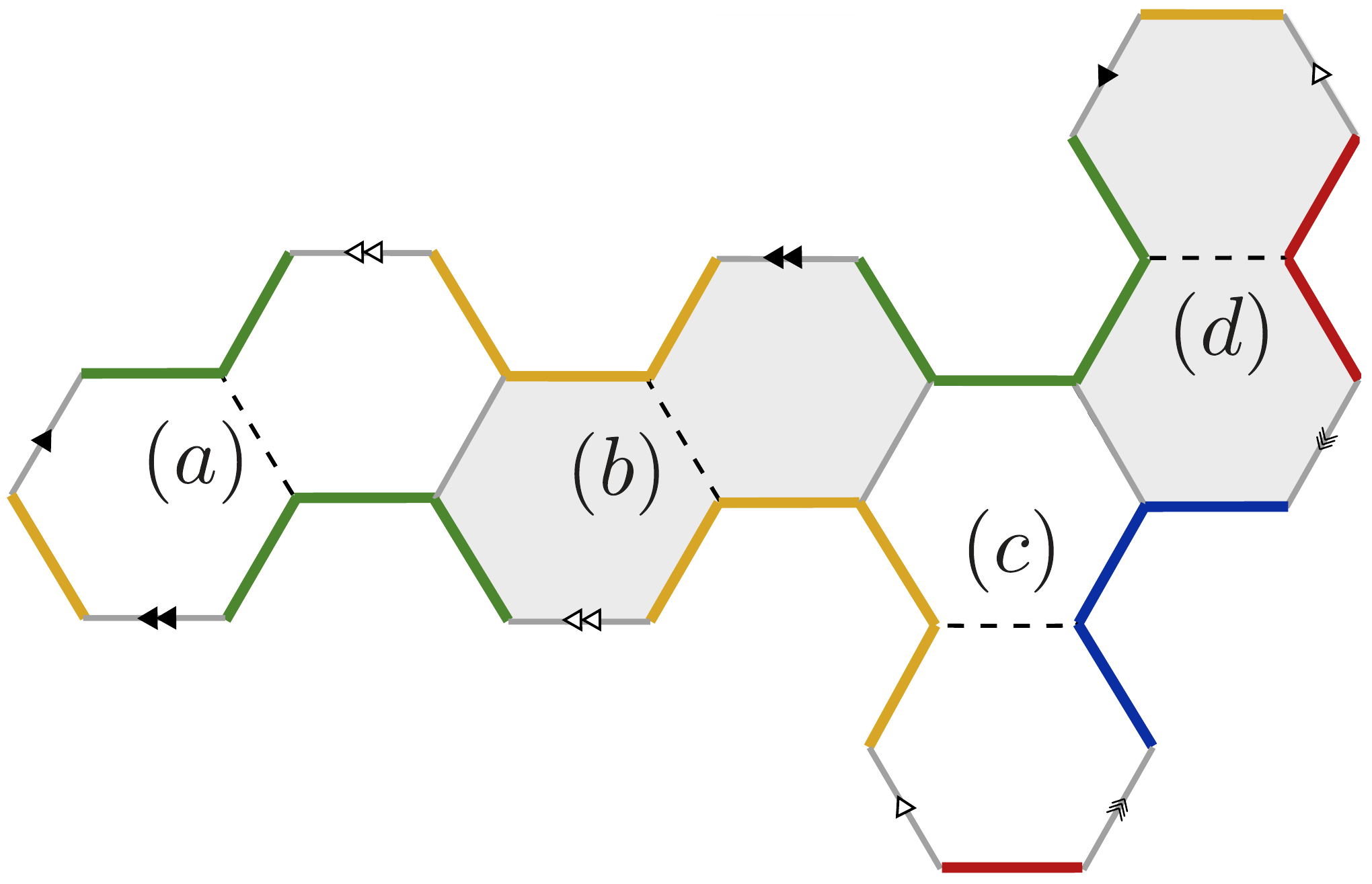}
\caption{The four octagons of case A.\label{fig:CaseFcheck}}
\end{figure}

Consider for illustration the particular case~A in~\figref{fig:3steps}. There are four octagons to be
considered, as shown in~\figref{fig:CaseFcheck}.
The first two contain pairs of physical edges
associated to the same external operator and thus give a vanishing
contribution, as can be easily seen by taking the limit
of two coinciding points for a generic octagon.
For each of the labelings in~\tabref{tab:labeltab},
the resulting expressions for the remaining two octagons are
summarized in~\tabref{tab:caseB}.
Accounting for the labeling and combinatorial factors listed
in~\tabref{tab:labeltab}, we can then read off the coefficient of
$X^{m} Y^{k-m}$ as
\begin{align}
\nonumber
&\text{case A}\rvert_{X^{m} Y^{k-m} \, \text{coeff.}} =k^4 \Big[ \frac{(r+1/2)^{4}+(r-1/2)^{4}}{24}  \times\\
& \left( 4 \mathcal{M} +2 \mathcal{M}^{2} \right)\!\! \sum_{a=-2}^{2}  X^{m+a} Y^{k-m-a}  \Big]_{X^{m} Y^{k-m}\,\text{coeff.}}
\nn
\end{align}
where we have used that $\mathcal{M}^{(1)}(1/z) = \mathcal{M}^{(1)}(z) \equiv \mathcal{M}
$. As explained above, this coefficient receives contributions from a
few neighboring polygonizations, accounted for by the sum in the
second line.
The remaining cases follow in complete analogy. When we sum them all,
we obtain a perfect match with~\eqref{eq:Fkm2largek}.

\begin{table}
\caption{Contributions for the one-particle octagons for each distinct
operator labeling of case A.\label{tab:caseB}}
\begin{ruledtabular}
\begin{tabular}{ccc}
{\footnotesize Labeling}
& {\footnotesize Octagon $(c)$}
& {\footnotesize Octagon $(d)$} \\
\colrule
$1234$ & $\mathcal{M}^{(1)}(z)$   & $\mathcal{M}^{(1)}(z)$   \\
$1324$ & $\mathcal{M}^{(1)}(1/z)$ & $\mathcal{M}^{(1)}(1/z)$ \\
$2413$ & $\mathcal{M}^{(1)}(1/z)$ & $\mathcal{M}^{(1)}(1/z)$ \\
$3412$ & $\mathcal{M}^{(1)}(z)$   & $\mathcal{M}^{(1)}(z)$   \\
\end{tabular}
\end{ruledtabular}
\end{table}

\section{Conclusions and Overlook}

We proposed here a novel formalism for computing correlation
functions of local gauge-invariant operators in $\mathcal{N}=4$ SYM
theory at any genus and any order in the coupling in the large $\Nc$
't Hooft expansion.

In this Letter, we already performed one very non-trivial check of our conjecture. We reproduced
the first non-planar correction to the correlation
function of four large BPS operators at one loop and two loops
from integrability. At the
end of the day, this computation is rather simple, and uses only
formulas for a single mirror particle already worked out in~\cite{Fleury:2016ykk}.
In an upcoming paper~\cite{Bargheer:2018jvq}, we perform numerous other checks
that probe all steps in our proposal in great detail:
The polygonization, the hexagonalization, the sprinkling, and the
stratification. These include finite-size corrections to the
computation above, correlators at strict finite size,
higher-genus examples, and subtleties related to the choice of the gauge group.
Through the operator product expansion of the obtained correlators, we can read off
conformal data of non-BPS operators beyond the planar limit.

One of the advantages of dealing with BPS external
operators (as considered in this Letter) is avoiding the subtlety of double-trace mixing.
It would be interesting to study the mixing effects.
(See~\cite{Eden:2017ozn} for very interesting first explorations in
this direction.)
It would also be important to better understand the integrand one
obtains after sprinkling the hexagons with a few mirror particles.
As we increase the number of mirror particles, it quickly becomes
monstrous. How do we tame it efficiently? Another interesting
problem -- which can be realistically addressed only once we
progress with the former -- concerns going to strong coupling and making
contact with
the recent exciting developments on the bootstrap approach to
loop corrections in AdS~\cite{\AdSLoops}.
One can then explore various interesting questions such as the
emergence of bulk locality~\cite{Heemskerk:2009pn,Maldacena:2015iua}.
Will we find higher-genus subtleties in our  integrability-based formalism akin to the
complications with supermoduli integrations recently observed in the Ramond--Neveu--Schwarz formalism~\cite{Witten:2012bg,Witten:2012ga,Witten:2012bh}?

Finally, a fun project would be to re-sum
the 't Hooft expansion -- perhaps starting with some simplifying kinematic limits. What awaits us there, and what can we learn
about string (field) theory?

\pdfbookmark[1]{Acknowledgments}{acknowledgments}
\subsection*{Acknowledgments}

We thank Benjamin Basso for numerous enlightening discussions and for
collaboration at initial stages of this project.
We are grateful for numerous important discussions with N.~Berkovits,
F.~Coronado, J.~Gomis, N.~Gromov, and J.~Maldacena.
T.\,B. thanks DESY Hamburg for its support and hospitality
during all stages of this work, and Perimeter Institute for a very
fruitful visit.
The work of J.\,C. and T.\,F. was supported by the
People Programme (Marie Curie Actions) of the European Union's Seventh
Framework Programme FP7/2007-2013/ under REA Grant Agreement No 317089
(GATIS), by the European Research Council (Programme ``Ideas''
ERC-2012-AdG 320769 AdS-CFT-solvable), from the ANR grant StrongInt
(BLANC-SIMI-4-2011).
The work of J.\,C. is supported by research Grant No.\ CERN/FIS-NUC/0045/2015.
T.\,F. thanks the warm hospitality of the Perimeter Institute and of
the Laboratoire de Physique Th\'eorique de l'Ecole Normale
Sup\'erieure, where a large part of this work was done. T.\,F. thanks
also CAPES (Coordenação de Aperfeiçoamento de Pessoal de Nível
Superior) grant INCTMAT process 88887.143256/2017-0 and the
Perimeter Institute for financial support.
S.\,K. acknowledges support from the Institute for Advanced Study.
Research at the Perimeter Institute is supported in part by the
Government of Canada through NSERC and by the Province
of Ontario
through MRI.
This research received funding from the Simons Foundation
Grant No.\ 488661 (Simons collaboration on the non-perturbative bootstrap)
and FAPESP Grant No.\ \mbox{2016/01343-7}.

\pdfbookmark[1]{\refname}{references}
\bibliography{references}

\end{document}